\documentclass[amsmath,aps,showpacs,a4paper,10pt]{revtex4}

 \usepackage{epsf}
 \usepackage{graphicx}    

 \usepackage{enumerate}   

\begin{document}

 \newcommand{\be}[1]{\begin{equation}\label{#1}}
 \newcommand{\ee}{\end{equation}}
 \newcommand{\bea}{\begin{eqnarray}}
 \newcommand{\eea}{\end{eqnarray}}
 \def\disp{\displaystyle}

 \def\gsim{ \lower .75ex \hbox{$\sim$} \llap{\raise .27ex \hbox{$>$}} }
 \def\lsim{ \lower .75ex \hbox{$\sim$} \llap{\raise .27ex \hbox{$<$}} }

 \begin{titlepage}

 \begin{flushright}
 arXiv:1902.00289
 \end{flushright}

 \title{\Large \bf Observational Constraints on Growth Index\\
 with Cosmography}

 \author{Zhao-Yu~Yin\,}
 \email[\,email address:\ ]{854587902@qq.com}
 \affiliation{School of Physics,
 Beijing Institute of Technology, Beijing 100081, China}

 \author{Hao~Wei\,}
 \email[\,Corresponding author;\ email address:\ ]{haowei@bit.edu.cn}
 \affiliation{School of Physics,
 Beijing Institute of Technology, Beijing 100081, China}

 \begin{abstract}\vspace{1cm}
 \centerline{\bf ABSTRACT}\vspace{2mm}
 In the literature, it was proposed that the growth index
 $\gamma$ is useful to distinguish the scenarios of dark energy
 and modified gravity. In the present work, we consider the
 constraints on the growth index $\gamma$ by using the latest
 observational data. To be model-independent, we
 use cosmography to describe the cosmic expansion history, and
 also expand the general $\gamma(z)$ as a Taylor series with
 respect to redshift $z$ or $y$-shift, $y=z/(1+z)$. We find
 that the present value $\gamma_0=\gamma(z=0)\simeq 0.42$ (for
 most of viable $f(R)$ theories) is inconsistent with the
 latest observational data at high confidence level (C.L.). On
 the other hand, $\gamma_0\simeq 0.55$ (for dark energy models
 in GR) can be consistent with the latest observational data at
 $1\sigma$ C.L. in 5 of the 9 cases under consideration, but is
 inconsistent beyond $2\sigma$ C.L. in the other 4 cases (while
 it is still consistent within the $3\sigma$ region). Thus, we
 can say nothing firmly about $\gamma_0\simeq 0.55$. We also
 find that a varying $\gamma(z)$ is favored.
 \end{abstract}


 \maketitle

 \end{titlepage}

 \renewcommand{\baselinestretch}{1.0}


\section{Introduction}\label{sec1}

It is a great mystery since the current accelerated expansion
 of our universe was discovered in 1998~\cite{Riess:1998cb,
 Perlmutter:1998np}. More than 20 years passed, and we still do
 not know the very nature of the cosmic acceleration by now.
 Usually, an unknown energy component with negative pressure
 (dark energy) is introduced to interpret this mysterious
 phenomenon in general relativity (GR). Alternatively, one can
 make a modification to GR (modified gravity). In fact,
 modified gravity can also successfully explain the cosmic
 acceleration without invoking dark energy. So far, these two
 scenarios are both competent~\cite{Ishak:2018his,Amendola:2016saw,
 Joyce:2014kja,Clifton:2011jh}.

In order to understand the nature of the cosmic accelerated
 expansion, one of the most important tasks is to distinguish
 between the scenarios of dark energy and modified gravity. If
 the observational data could help us to confirm or exclude
 one of these two scenarios as the real cause of
 this mysterious phenomenon, it will be a great step forward.
 However, many cosmological observations merely probe the
 cosmic expansion history. Unfortunately, as is well known (see
 e.g.~\cite{Sahni:2006pa}), one can always build models sharing
 a same cosmic expansion history, and hence these models cannot
 be distinguished by using the observational data of the
 expansion history only. So, some independent and complementary
 probes are required. Later, it is proposed that if the
 cosmological models share a same cosmic expansion history,
 they might have different growth histories, which are characterized
 by the matter density contrast $\delta(z)\equiv\delta\rho_m/\rho_m$
 as a function of redshift $z$. Therefore, they might be
 distinguished from each other by combining the observations of
 both the expansion and growth histories (see
 e.g.~\cite{Linder:2005in,Linder:2007hg,Huterer:2006mva,
 Wang:2007ht,Zhang:2007nk,Jain:2007yk,Wei:2008ig,
 Wei:2008vw,Wei:2013rea,Yin:2018mvu,Viznyuk:2018eiz,Amendola:2014yca,
 Basilakos:2016nyg} and references therein).

It is convenient to introduce the growth rate
 $f\equiv d\ln\delta/d\ln a$, where $a=(1+z)^{-1}$ is the scale
 factor. As is well known, a good parameterization for the
 growth rate is given
 by~\cite{Peebles1980,Lahav:1991wc,Wang:1998gt,Lue:2004rj}
 \be{eq1}
 f\equiv\frac{d\ln\delta}{d\ln a}=\Omega_m^\gamma\,,
 \ee
 where $\gamma$ is the growth index, and $\Omega_m$ is the
 fractional energy density of matter. Beginning
 in e.g.~\cite{Linder:2005in,Linder:2007hg}, it was advocated
 that the growth index $\gamma$ is useful to distinguish the
 scenarios of dark energy and modified gravity. For example,
 it is found that $\gamma=6/11\simeq 0.545$ for $\Lambda$CDM
 model~\cite{Linder:2005in,Linder:2007hg}, and
 $\gamma\simeq 0.55$ for most of dark energy
 models in GR~\cite{Linder:2005in}. In fact, they are clearly
 distinct from the ones of modified gravity theories. For
 instance, it is found that $\gamma\simeq 0.68$
 for Dvali-Gabadadze-Porrati (DGP) braneworld
 model~\cite{Linder:2007hg,Wei:2008ig}, and $\gamma\simeq 0.42$
 for most of viable $f(R)$ theories~\cite{Gannouji:2008wt,
 Tsujikawa:2009ku,Shafieloo:2012ms,Tsujikawa:2010zza}. In
 general, the growth index $\gamma$ is a function
 of redshift~$z$. It is argued that $\gamma(z)$ lies in a relatively
 narrow range around the above values respectively, and hence
 one might distinguish between them.

In the literature, most of the relevant works assumed a
 particular cosmological model to obtain the growth index $\gamma$.
 Thus, the corresponding results are model-dependent in fact.
 However, robust results should be model-independent. So, it is
 of interest to obtain the growth index $\gamma$ from the
 observational data by using a model-independent approach. In
 fact, recently we have made an effort in~\cite{Yin:2018mvu} to
 obtain a non-parametric reconstruction of the growth index $\gamma$
 via Gaussian processes by using the latest observational data.
 Although the approach of Gaussian processes is clearly
 model-independent, its reliability at high redshift might be
 questionable. So, it is of interest to test the growth index
 $\gamma$ by using a different method, and cross-check the
 corresponding results with the ones from Gaussian processes.

As is well known, one of the powerful model-independent
 approaches is cosmography~\cite{Weinberg1972,Visser:2003vq,
 Bamba:2012cp,Cattoen:2008th,Vitagliano:2009et,Cattoen:2007id,
 Xu:2010hq,Xia:2011iv,Zhang:2016urt,Dunsby:2015ers,Luongoworks,
 Zhou:2016nik,Zou:2017ksd,Luongo:2013rba}. In fact, the only
 necessary assumption of cosmography is the cosmological
 principle. With cosmography, one can analyze the evolution of
 the universe without assuming any particular cosmological
 model. Essentially, cosmography is the Taylor series expansion
 of the quantities related to the cosmic expansion history
 (especially the luminosity distance $d_L$), and hence it is
 model-independent indeed. In the present work, we will
 constrain the growth index $\gamma$ by using the latest
 observational data via the cosmographic approach. However,
 there are several shortcomings in the usual cosmography (see
 e.g.~\cite{Zhou:2016nik}). For instance, it is plagued with
 the problem of divergence or an unacceptably large error, and
 it fails to predict the future evolution of the universe.
 Thus, some generalizations of cosmography inspired by the
 Pad\'e approximant were proposed in~\cite{Zhou:2016nik} (see
 also e.g.~\cite{Gruber:2013wua,Wei:2013jya,Liu:2014vda,
 Adachi:2011vu,Capozziello:2018jya,Aviles:2014rma,Capozziello:2017ddd,
 Capozziello:2017nbu,Capozziello:2018aba}), which can avoid or
 at least alleviate the problems of ordinary cosmography. So,
 we also consider the Pad\'e cosmography in this work.

The rest of this paper is organized as follows.
 In Sec.~\ref{sec2}, we describe the methodology to constrain
 the growth index $\gamma$ by using the latest observational
 data. In Secs.~\ref{sec3} and \ref{sec4}, we obtain the
 corresponding constraints on $\gamma$ with the
 $z$-cosmography, the $y$-cosmography, and the
 Pad\'e cosmography, respectively.
 In Sec.~\ref{sec5}, conclusion and discussion are given.


\section{Methodology}\label{sec2}

In the literature, there are many approaches to deal with the
 growth history. For example, one can consider a Lagrangian
 derived from an effective field theory (EFT)
 expansion~\cite{Gubitosi:2012hu,Gleyzes:2013ooa} (see also
 e.g.~\cite{Ade:2015rim}), and implement the full background
 and perturbation equations for this action in the Boltzmann
 code EFTCAMB/EFTCosmoMC~\cite{Hu:2013twa,Raveri:2014cka,Hu:2014oga}.
 The second approach is more
 phenomenological~\cite{Zhao:2008bn,Zhao:2010dz,Simpson:2012ra,
 Macaulay:2013swa,Baker:2014zva,Daniel:2010ky,Hojjati:2011ix}
 (see also e.g.~\cite{Ade:2015rim,Xu:2013tsa}), by directly
 parameterizing the functions of the gravitational potentials
 $\Phi$ and $\Psi$, such as $\mu=G_{\rm eff}/G$,
 $\eta=\Phi/\Psi$, and/or $\Sigma$, $Q$, in the modified
 relativistic Poisson equations. It can be implemented by using the
 code MGCAMB~\cite{Zhao:2008bn,Hojjati:2011ix} integrated in
 CosmoMC~\cite{Lewis:2002ah}. The third approach is the
 simplest one, by directly parameterizing the growth rate $f$
 as in Eq.~(\ref{eq1}), with no need for numerically solving
 the perturbation equations. For simplicity, we choose this
 approach in the present work.

By definition $f\equiv d\ln\delta/d\ln a$, it is easy to get
 (see e.g.~\cite{Linder:2009kq,Yin:2018mvu,Wang:2012fq})
 \be{eq2}
 \frac{\delta}{\delta_0}
 =\exp\left(\int_1^a \frac{f\,d\tilde{a}}{\tilde{a}}\right)=
 \exp\left(-\int_0^z\frac{f\,d\tilde{z}}{1+\tilde{z}}\right)\,,
 \ee
 where the subscript ``\,0\,'' indicates the present value of
 the corresponding quantity, namely $\delta_0=\delta(z=0)$. On
 the other hand, the cosmic expansion history can
 be characterized by the luminosity distance
 $d_L=\left(c/H_0\right)D_L$, where $c$ is the speed of light,
 $H_0$ is the Hubble constant, and (see e.g. the
 textbooks~\cite{Weinberg1972})
 \be{eq3}
 D_L\equiv(1+z)\int_0^z\frac{d\tilde{z}}{E(\tilde{z})}\,,
 \ee
 in which $E\equiv H/H_0$, and the Hubble
 parameter $H\equiv\dot{a}/a$ (where a dot denotes the
 derivative with respect to cosmic time $t$). Note that we
 consider a flat Friedmann-Robertson-Walker (FRW) universe in
 this work. As is well known, $E(z)$ is free of $H_0$ actually.
 Differentiating Eq.~(\ref{eq3}), we get~\cite{Zou:2017ksd}
 \be{eq4}
 \frac{1+z}{E(z)} = \frac{d D_L}{dz}-\frac{D_L}{1+z} \,.
 \ee
 If the luminosity distance $d_L$ (or equivalently $D_L$) is
 known (in fact it will be given by the cosmography as below),
 we can obtain the dimensionless Hubble parameter $E(z)$ by
 using Eq.~(\ref{eq4}). Then, the fractional energy density of
 matter is given by
 \be{eq5}
 \Omega_m(z)\equiv\frac{8\pi G\rho_m}{3H^2}=
 \frac{\Omega_{m0}(1+z)^3}{E^2(z)}\,.
 \ee
 So, the growth rate $f=\Omega_m^\gamma$ is on hand, and hence
 $\delta/\delta_0$ in Eq.~(\ref{eq2}) is ready.

The data of the growth rate $f$ can be obtained from redshift space
 distortion (RSD) measurements. In fact, the observational $f_{obs}$
 data have been used in some relevant works (e.g.~\cite{Wei:2008ig,
 Gonzalez:2016lur,Gonzalez:2017tcm}). However, it is sensitive
 to the bias parameter $b$ which can vary in the range
 $b\in[1,\,3]$. This makes the observational $f_{obs}$ data
 unreliable~\cite{Nesseris:2017vor}. Instead, the combination
 $f\sigma_8(z)\equiv f(z)\,\sigma_8(z)$ is independent of the
 bias, and hence is more reliable, where $\sigma_8(z)=\sigma_8(z=0)
 \,\delta(z)/\delta_0=\sigma_{8,\,0}\,\delta(z)/\delta_0$ is
 the redshift-dependent rms fluctuations of the linear density
 field within spheres of radius $8h^{-1}$Mpc~\cite{Nesseris:2017vor}.
 In fact, the observational $f\sigma_{8,\,obs}$ data can be obtained
 from weak lensing or RSD measurements~\cite{Nesseris:2017vor,
 Kazantzidis:2018rnb}. In the present work, we use the sample
 consisting of 63 observational $f\sigma_{8,\,obs}$ data
 published in~\cite{Kazantzidis:2018rnb}, which is the largest
 $f\sigma_8$ compilation in the literature by now. As mentioned
 above, once $D_L$ is given, we can get the theoretical $f\sigma_8$
 by using Eqs.~(\ref{eq4}), (\ref{eq5}), and (\ref{eq1}),
 (\ref{eq2}). Thus, the $\chi^2$ from the $f\sigma_8$ data is
 given by
 \be{eq6}
 \chi^2_{f\sigma_8}=\sum_i \frac{\left[\,f\sigma_{8,\,obs}(z_i)
 -f\sigma_{8,\,{\rm mod}}(z_i)\,\right]^2}
 {\sigma_{f\sigma_8}^2 (z_i)}\,.
 \ee

It is easy to see that only using the observational $f\sigma_8$ data
 is not enough to constrain the model parameter $\Omega_{m0}$,
 and the cosmographic parameters $q_0$, $j_0$ ... in $D_L$.
 Since they mainly affect the cosmic expansion history, we also
 use such kinds of observations. Obviously, the type Ia
 supernovae (SNIa) data is useful. Here, we consider the
 Pantheon sample~\cite{Scolnic:2017caz,Pantheondata,Pantheonplugin}
 consisting of 1048 SNIa, which is
 the largest spectroscopically confirmed SNIa sample by
 now. The corresponding $\chi^2$ is given by
 \be{eq7}
 \chi^2_{\rm Pan}=\Delta\boldsymbol{m}^{\,T}\cdot
 \boldsymbol{C}^{-1}\cdot\Delta\boldsymbol{m}\,,
 \ee
 where for the $i$-th SNIa, $\Delta m_i=m_i-m_{{\rm mod},\,i}\,$, and
 $\boldsymbol{C}$ is the total covariance matrix,
 \be{eq8}
 m_{\rm mod} = 5\log_{10} D_L + {\cal M}\,,
 \ee
 in which $\cal M$ is a nuisance parameter corresponding to
 some combination of the absolute magnitude $M$ and $H_0$. We
 refer to~\cite{Scolnic:2017caz,Pantheondata,Pantheonplugin}
 for technical details (see also e.g.~\cite{Deng:2018jrp}).
 Since $H_0$ is absorbed into $\cal M$ in the analytic
 marginalization, the Pantheon SNIa sample is free of the
 Hubble constant $H_0$.

We further consider the observational data from the baryon acoustic
 oscillation (BAO). Note that there exist many kinds of BAO data in
 the literature, such as $D_V(z)$, $d_z\equiv r_s(z_d)/D_V(z)$,
 $D_A(z)/r_d$, $D_M(z)/r_d$, $H(z)\,r_s(z_d)$ and $A$. However,
 in the former ones, they will introduce one or more extra
 model parameters, for instance $H_0$, and/or $\Omega_b h^2$. Since
 the $f\sigma_8$ data, the SNIa data, the cosmography for $D_L$, and
 other data are all free of $H_0$ and $\Omega_b h^2$, we choose
 to avoid introducing extra model parameters here. Thus, in
 this work, we use the BAO data only in the form of (see
 e.g.~\cite{Eisenstein:2005su,Blake:2011en})
 \be{eq9}
 A\equiv\Omega_{m0}^{1/2}H_0 D_V/(cz)
 =\frac{\Omega_{m0}^{1/2}}{z}\left[\frac{D_L^2}{(1+z)^2}\cdot
 \frac{z}{E(z)}\,\right]^{1/3}\,,
 \ee
 which does not introduce extra model parameters since the
 factor $c/H_0$ in $D_V$ is canceled. We consider the six data
 of the acoustic parameter $A(z)$ compiled in the last column
 of Table~3 of~\cite{Blake:2011en}. The first data point from
 6dFGS is uncorrelated with other five ones, and hence its
 $\chi^2_{\rm 6dFGS}=(A_{obs}-A_{\rm mod})^2/\sigma^2$
 directly. The 2nd and 3rd data points from SDSS are correlated
 with coefficient $0.337$, and hence the inverse covariance
 matrix of these two data points is given by
 \be{eq10}
 \boldsymbol{C}^{-1}_{\rm SDSS}=
 \left(
  \begin{array}{cc}
    4406.72 & -1485.06 \\
    -1485.06 & 4406.72
  \end{array}
 \right)\,.
 \ee
 The inverse covariance matrix of the last three data points
 from WiggleZ is given in Table~2 of~\cite{Blake:2011en},
 \be{eq11}
 \boldsymbol{C}^{-1}_{\rm WiggleZ}=
 \left(
  \begin{array}{ccc}
    1040.3 & -807.5 & 336.8 \\
    -807.5 & 3720.3 & -1551.9 \\
    336.8 & -1551.9 & 2914.9
  \end{array}
 \right)\,.
 \ee
 The $\chi^2$ from the data of SDSS and WiggleZ are both given
 in the form of $\chi^2=\Delta\boldsymbol{A}^{\,T}\cdot
 \boldsymbol{C}^{-1}\cdot\Delta\boldsymbol{A}$. Thus, the total
 $\chi^2$ from the BAO data is $\chi^2_{\rm BAO}=
 \chi^2_{\rm 6dFGS}+\chi^2_{\rm SDSS}+\chi^2_{\rm WiggleZ}$.

On the other hand, the free parameter $\sigma_{8,\,0}$ cannot
 be well constrained by using the $f\sigma_8$ data and the
 observations of the expansion history. Fortunately, in
 the literature there are many observational data of the combination
 $S_8\equiv\sigma_{8,\,0}(\Omega_{m0}/0.3)^{0.5}$ from the
 cosmic shear observations~\cite{Kilbinger:2014cea}, which can
 be used to constrain both the free parameters $\sigma_{8,\,0}$
 and $\Omega_{m0}$. Here, we consider the ten data points given
 in Table~\ref{tab1}. The corresponding $\chi^2_{S_8}=
 \sum_i \, (S_{8,\,obs,\,i}-S_{8,\,{\rm mod},\,i})^2 /
 \sigma_{S_8,\,i}^2\,$. Note that if the upper and the lower
 uncertainties of the data are not equal, we choose the bigger
 one as $\sigma_{S_8,\,i}$ conservatively.

In fact, there are other kinds of observational data in the
 literature. However, we do not use them here, to avoid introducing
 extra model parameters, as mentioned above. For instance, if
 we want to use the 51 observation $H(z)$ data compiled
 in~\cite{Magana:2017nfs} (the largest sample by now to our
 best knowledge), an extra free parameter $H_0$ is necessary.
 So, we give up. On the other hand, since the usual cosmography
 cannot work well at very high redshift, we also do not consider the
 observational data from cosmic microwave background (CMB) at
 redshift $z\sim 1090$. Otherwise, the cosmographic parameters
 should be fine-tuned. However, the Pad\'e cosmography works well at
 very high redshift, and hence we can use the CMB data in this
 case (see Sec.~\ref{sec4}).

All the model parameters can be constrained by using the
 observational data to perform a $\chi^2$ statistics. Here, the
 total $\chi^2_{\rm tot}=\chi^2_{f\sigma_8}+\chi^2_{\rm Pan}+
 \chi^2_{\rm BAO}+\chi^2_{S_8}$. In the following, we use the
 Markov Chain Monte Carlo (MCMC) code CosmoMC~\cite{Lewis:2002ah} to
 this end.


\begin{table}[tb]
 \renewcommand{\arraystretch}{1.6}
 \begin{center}
 \vspace{-3mm}  
 \begin{tabular}{rcl|rcl} \hline\hline
  Survey &  $S_8$  &  Ref.\ \ \ \ & Survey &  $S_8$  &  Ref. \\ \hline
  HSC & \ \ \ \ $0.780^{+0.030}_{-0.033}$ \ \ \ \ & \cite{Hikage:2018qbn} & DES (c.s.) & \ \ \ \ $0.782^{+0.027}_{-0.027}$ \ \ \ \ & \cite{Troxel:2017xyo} \\
  DES (g.c.+\,w.l.) & $0.773^{+0.026}_{-0.020}$  & \cite{Abbott:2017wau} &  CFHTLenS &  $0.732^{+0.029}_{-0.031}$  & \cite{Joudaki:2016mvz,Hikage:2018qbn} \\
  KiDS-450 (c.f.) & $0.745\pm 0.039$ & \cite{Hildebrandt:2016iqg} &  KiDS-450 (p.s.) & $0.651\pm 0.058$ & \cite{Kohlinger:2017sxk} \\
  DLS & $0.818^{+0.034}_{-0.026}$ & \cite{Jee:2015jta} & KiDS-450 + GAMA & $0.800^{+0.029}_{-0.027}$ & \cite{vanUitert:2017ieu} \\
  KiDS-450 + 2dFLenS & $0.742\pm 0.035$ & \cite{Joudaki:2017zdt} & \ \ \ Planck 2018 CMB lensing &  $0.832\pm 0.013$  & \cite{Aghanim:2018eyx} \\
 \hline\hline
 \end{tabular}
 \end{center}
 \caption{\label{tab1} The observational data of
 $S_8\equiv\sigma_{8,\,0}(\Omega_{m0}/0.3)^{0.5}$. See the
 Refs. for details.}
 \end{table}



\begin{table}[tb]
 \renewcommand{\arraystretch}{1.6}
 \begin{center}
 \vspace{3mm}  
 \begin{tabular}{c|c|c|c} \hline\hline
  Parameters \ &  Case $z$-0  &  Case $z$-1 & Case $z$-2 \\ \hline
  $\Omega_{m0}$ & \ $0.2858_{-0.0122 }^{+0.0121}{}_{-0.0236}^{+0.0244}{}_{-0.0310}^{+0.0325}$ \
     & \ $0.2983_{-0.0136 }^{+0.0136}{}_{-0.0274}^{+0.0268}{}_{-0.0364}^{+0.0358}$ \
     & \ \ $0.2887_{-0.0126 }^{+0.0126}{}_{-0.0248}^{+0.0254}{}_{-0.0324}^{+0.0340}$  \\
  $q_0$ & \ \ $-0.5441_{-0.0287 }^{+0.0287}{}_{-0.0558}^{+0.0570}{}_{-0.0730}^{+0.0751}$ \ \
     & \ \ $-0.4803_{-0.0378 }^{+0.0475}{}_{-0.0932}^{+0.0872}{}_{-0.1368}^{+0.1063}$ \ \
     & \ \ $-0.5260_{-0.0387 }^{+0.0386}{}_{-0.0747}^{+0.0772}{}_{-0.0971}^{+0.1039}$  \\
  $j_0$ & \ $0.6339_{-0.0667 }^{+0.0668}{}_{-0.1323}^{+0.1315}{}_{-0.1735}^{+0.1725}$ \
     & \ $0.6079_{-0.0603 }^{+0.0600}{}_{-0.1187}^{+0.1234}{}_{-0.1556}^{+0.1695}$ \
     & \ \ $0.6435_{-0.0660 }^{+0.0662}{}_{-0.1293}^{+0.1328}{}_{-0.1692}^{+0.1768}$  \\
  $\sigma_{8,\,0}$ & \ $0.8109_{-0.0189 }^{+0.0189}{}_{-0.0363}^{+0.0385}{}_{-0.0474}^{+0.0516}$ \
     & \ $0.7924_{-0.0215 }^{+0.0181}{}_{-0.0370}^{+0.0418}{}_{-0.0475}^{+0.0573}$ \
     & \ \ $0.8099_{-0.0206 }^{+0.0189}{}_{-0.0377}^{+0.0403}{}_{-0.0493}^{+0.0539}$  \\
  $\gamma_{0}$ & \ $0.6281_{-0.0389 }^{+0.0387}{}_{-0.0752}^{+0.0786}{}_{-0.0983}^{+0.1044}$ \
     & \ $0.6679_{-0.0459 }^{+0.0462}{}_{-0.0915}^{+0.0921}{}_{-0.1227}^{+0.1223}$ \
     & \ \ $0.5925_{-0.0495 }^{+0.0494}{}_{-0.0948}^{+0.1006}{}_{-0.1227}^{+0.1349}$  \\
  $\gamma_{1}$ & N/A & \ $-0.2676_{-0.1302 }^{+0.0364}{}_{-0.1843}^{+0.2803}{}_{-0.2130}^{+0.5265}$ \
     & \ \ $0.2786_{-0.2464 }^{+0.2222}{}_{-0.4439}^{+0.4765}{}_{-0.5792}^{+0.6400}$  \\
  $\gamma_{2}$ & N/A & N/A & \ \ $-0.3003_{-0.1120 }^{+0.1300}{}_{-0.2533}^{+0.2268}{}_{-0.3448}^{+0.2962}$ \\
 \hline\hline
 \end{tabular}
 \end{center}
 \caption{\label{tab2} The mean with $1\sigma$, $2\sigma$,
 $3\sigma$ marginalized uncertainties of the model parameters
 for the cases with the $z$-cosmography and $\gamma=\gamma_0$
 (labeled as ``\,$z$-0\,''), $\gamma=\gamma_0+\gamma_1\,z$ (labeled
 as ``\,$z$-1\,''), $\gamma=\gamma_0+\gamma_1\,z+\gamma_2\,z^2$
 (labeled as ``\,$z$-2\,''). See the text for details.}
 \end{table}



 \begin{center}
 \begin{figure}[tb]
 \centering
 \vspace{-11mm}  
 \includegraphics[width=0.7\textwidth]{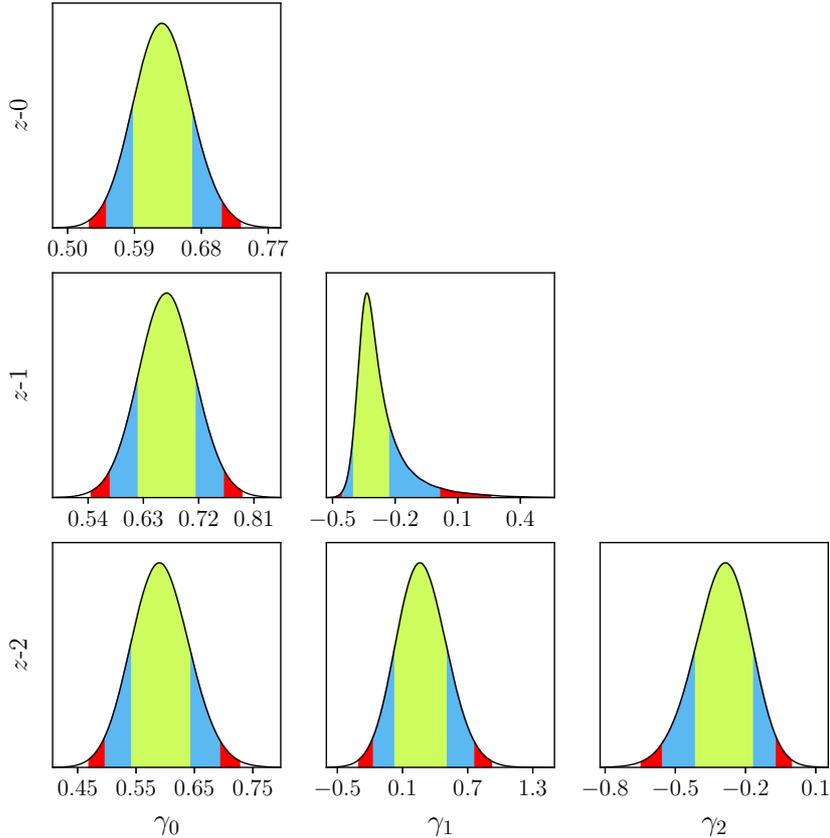}
 \caption{\label{fig1} The 1D marginalized probability
 distributions of the parameters related to $\gamma$. The
 $1\sigma$, $2\sigma$, $3\sigma$ uncertainties are shown by
 the green, blue, red regions, respectively. The top, middle,
 bottom panels correspond to the $z$-0, $z$-1, $z$-2 cases,
 respectively. See the text and Table~\ref{tab2} for details.}
 \end{figure}
 \end{center}


\vspace{-12mm}  


\section{Observational constraints with the ordinary cosmography}\label{sec3}


\subsection{The case of $z$-cosmography}\label{sec3a}

At first, we consider the case of $z$-cosmography. Introducing
 the so-called cosmographic parameters, namely the Hubble
 constant $H_0$, the deceleration $q_0$, the jerk $j_0$, the
 snap $s_0$, ...
 \be{eq12}
 H_0\equiv\left.\frac{1}{a}\frac{da}{dt}\right|_{t=t_0}\,,\quad
 q_0\equiv\left.-\frac{1}{aH^2}\frac{d^{2}a}{dt^{2}}\right|_{t=t_0}\,,\quad
 j_0\equiv\left.\frac{1}{aH^3}\frac{d^{3}a}{dt^{3}}\right|_{t=t_0}\,,\quad
 s_0\equiv\left.\frac{1}{aH^4}\frac{d^{4}a}{dt^{4}}\right|_{t=t_0}\,,\quad ...
 \ee
 one can express the quantities related to the cosmic expansion
 history, e.g. the scale factor $a(t)$, the Hubble parameter
 $H(z)$, and the luminosity distance $d_L(z)$, as a Taylor
 series expansion (see e.g.~\cite{Weinberg1972,Visser:2003vq,
 Bamba:2012cp,Cattoen:2008th,Vitagliano:2009et,Cattoen:2007id,
 Xu:2010hq,Xia:2011iv,Zhang:2016urt,Dunsby:2015ers,Luongoworks,
 Zhou:2016nik,Zou:2017ksd} and references therein). The most
 important one is the luminosity distance $d_L(z)$, and its
 Taylor series expansion with respect to redshift $z$ reads
 (see e.g.~\cite{Weinberg1972,Visser:2003vq,Bamba:2012cp,
 Zhou:2016nik,Zou:2017ksd} for
 details)
 \bea
 d_L(z)&=&\frac{cz}{H_0}\left[1+\frac{1}{2}\left(1-q_0\right)z
 -\frac{1}{6}\left(1-q_0-3q_0^2+j_0\right)z^2\right.\nonumber\\[1mm]
 &&\left.+\frac{1}{24}\left(2-2q_0-15q_0^2-15q_0^3+5j_0
 +10q_0 j_0+s_0\right)z^3+{\cal O}(z^4)\right]\,.\label{eq13}
 \eea
 Since the constraints become loose if the number of free
 parameters increases, we only consider the cosmography up to
 third order. Thus, the dimensionless luminosity
 distance $D_L=H_0 d_L/c$ is given by
 \be{eq14}
 D_L(z)=z+\frac{1}{2}\left(1-q_0\right)z^2
 -\frac{1}{6}\left(1-q_0-3q_0^2+j_0\right)z^3+{\cal O}(z^4)\,,
 \ee
 in which only two free cosmographic parameters $q_0$ and $j_0$
 are involved. Note that the Hubble constant $H_0$ does not
 appear, since the factor $c/H_0$ in $d_L$ is canceled.

In the literature, the growth index $\gamma$ is often assumed
 to be constant (see e.g.~\cite{Peebles1980,Xu:2013tsa,Zhang:2014lfa,
 Zhao:2017jma}). However, in general it is varying as a function of
 redshift $z$. To be model-independent, we can also expand
 $\gamma(z)$ as a Taylor series with respect to redshift $z$,
 namely $\gamma(z)=\gamma_0+\gamma_1\,z+\gamma_2\,z^2+...$,
 where the coefficients $\gamma_0$, $\gamma_1$, $\gamma_2$ ...
 are constants. Here, we consider three cases, labeled as
 ``\,$z$-0\,'', ``\,$z$-1\,'', ``\,$z$-2\,'', in which
 $\gamma(z)$ is Taylor expanded up to zeroth, first, second
 orders, respectively.

Substituting Eq.~(\ref{eq14}) into Eq.~(\ref{eq4}), we can get
 the dimensionless Hubble parameter $E(z)$. Using Eqs.~(\ref{eq5}),
 (\ref{eq1}), (\ref{eq2}), and $\gamma$, we
 obtain $f=\Omega_m^\gamma$ and then $f\sigma_8$. Substituting
 $D_L(z)$ and $E(z)$ into Eqs.~(\ref{eq8}) and (\ref{eq9}), we
 find $m_{\rm mod}$ and $A_{\rm mod}$. Finally, the total
 $\chi^2_{\rm tot}$ is ready.

By using the latest observational data, we obtain the constraints on
 all the model parameters involved, and present them in
 Table~\ref{tab2}, for the $z$-0, $z$-1, $z$-2 cases. Since we
 mainly concern the parameters related to the growth index $\gamma$,
 namely $\gamma_0$, $\gamma_1$ and $\gamma_2$, we also present
 their 1D marginalized probability distributions in Fig.~\ref{fig1}.
 Obviously, in all cases, $q_0<0$ and $j_0>0$ far beyond
 $3\sigma$ confidence level (C.L.), and these mean that today
 the universe is accelerating, while the acceleration is still
 increasing. From Tabel~\ref{tab2} and Fig.~\ref{fig1}, it is
 easy to see that for all cases, $\gamma_0\simeq 0.42$ is
 inconsistent with the latest observational data far beyond
 $3\sigma$ C.L. Note that $\gamma_0\simeq 0.55$ is consistent
 with the latest observational data within the $1\sigma$ region for
 the $z$-2 case, but is inconsistent beyond $2\sigma$ C.L. for both
 the $z$-0 and $z$-1 cases (while it is still consistent within the
 $3\sigma$ region). On the other hand, a varying $\gamma$ with
 non-zero $\gamma_1$ and/or $\gamma_2$ is favored. In the
 linear case with $\gamma=\gamma_0+\gamma_1\,z$ (namely the
 $z$-1 case), $\gamma_1<0$ in the $1\sigma$ region, and hence
 the growth index $\gamma$ decreases as redshift $z$ increases.
 In the quadratic case with $\gamma=\gamma_0+\gamma_1\,z+
 \gamma_2\,z^2$ (namely the $z$-2 case), $\gamma_2<0$ beyond
 $3\sigma$ C.L., and hence the function $\gamma(z)$ is a
 parabola opening down, namely $\gamma$ increases and then
 decreases as redshift $z$ increases. There exists an arched
 structure in the moderate redshift range. This result is quite
 similar to the one of~\cite{Yin:2018mvu}.


\subsection{The case of $y$-cosmography}\label{sec3b}

Let us turn to the case of $y$-cosmography. As is well known,
 a Taylor series with respect to redshift~$z$ converges
 only at low redshift $z$ around $0$, and it might diverge at
 high redshift $z>1$. In the literature (see
 e.g.~\cite{Cattoen:2008th,Cattoen:2007id,Vitagliano:2009et,
 Xia:2011iv,Zhou:2016nik,Zou:2017ksd}), a popular alternative
 to the $z$-cosmography is replacing $z$ with the so-called
 $y$-shift, $y=1-a=z/(1+z)$. Obviously, $y<1$ holds in the
 whole cosmic past $0\leq z<\infty$, and hence the Taylor
 series with respect to $y$-shift converges. In this case, we
 can expand the dimensionless luminosity distance
 $D_L=H_0 d_L/c$ as a Taylor
 series with respect to $y$ (see e.g.~\cite{Bamba:2012cp,
 Vitagliano:2009et,Zhou:2016nik,Zou:2017ksd} for details),
 \be{eq15}
 D_L(y)=y+\frac{1}{2}\left(3-q_0\right)y^2+
 \frac{1}{6}\left(11-5q_0+3q_0^2-j_0\right)y^3+{\cal O}(y^4)\,,
 \ee
 in which only two free cosmographic parameters $q_0$ and $j_0$
 are involved, since we only consider the cosmography up to
 third order in this work as mentioned above. Accordingly, here
 we also expand the growth index $\gamma$ as a Taylor series
 with respect to $y$, namely
 $\gamma(y)=\gamma_0+\gamma_1\,y+\gamma_2\,y^2+...$. Similarly,
 we consider three cases, labeled as
 ``\,$y$-0\,'', ``\,$y$-1\,'', ``\,$y$-2\,'', in which
 $\gamma(y)$ is Taylor expanded up to zeroth, first, second
 orders, respectively. Noting $y=z/(1+z)$ and
 $dF/dz=(1+z)^{-2}dF/dy$ for any function $F$, the formalism in
 Sec.~\ref{sec2} is still valid in the case of $y$-cosmography.

\begin{center}
 \begin{figure}[tb]
 \centering
 \vspace{-10mm}  
 \includegraphics[width=0.7\textwidth]{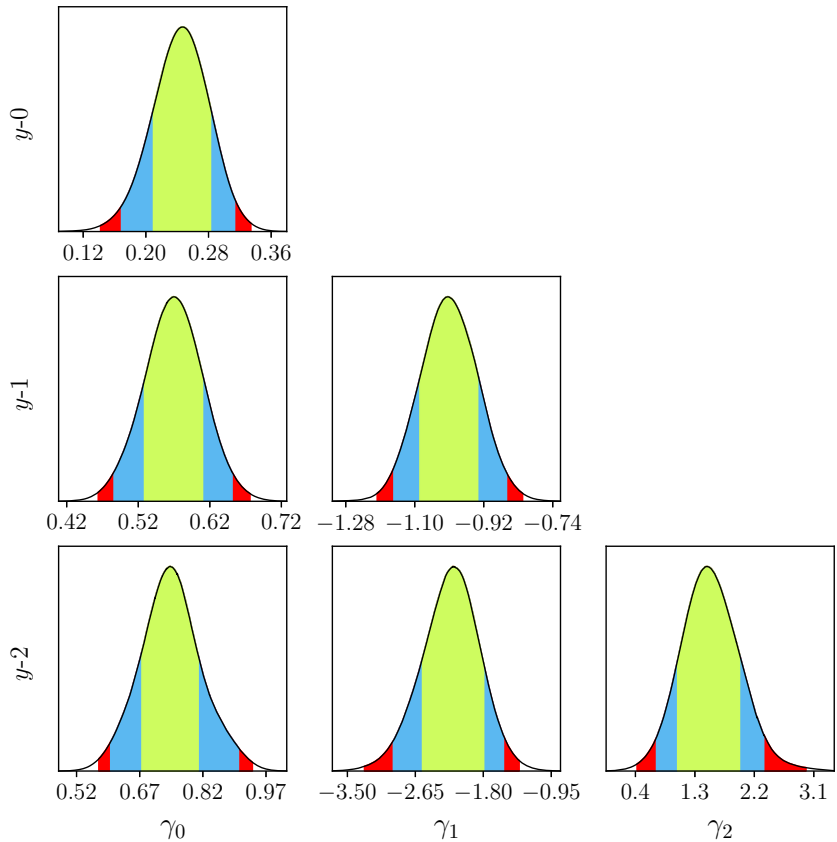}
 \caption{\label{fig2} The same as in Fig.~\ref{fig1}, except
 for the $y$-0, $y$-1, $y$-2 cases. See the text
 and Table~\ref{tab3} for details.}
 \end{figure}
 \end{center}

By using the latest observational data, we obtain
 the constraints on all the model parameters involved, and
 present them in Table~\ref{tab3}, for the $y$-0, $y$-1, $y$-2
 cases. In Fig.~\ref{fig2}, we also present the 1D marginalized
 probability distributions of the parameters related to the
 growth index $\gamma$, namely $\gamma_0$, $\gamma_1$ and
 $\gamma_2$. Obviously, the $y$-0 case is fairly different from
 the $y$-1, $y$-2 cases. In fact, $\gamma_0\,\lsim\,0.34$
 beyond $3\sigma$ C.L., and $q_0>0$ also beyond $3\sigma$ C.L.
 in the $y$-0 case. The unusual result that the universe is
 decelerating ($q_0>0$) suggests that the $y$-0 case with a
 constant $\gamma=\gamma_0$ is not competent to describe the
 real universe, and consequently $\gamma$ should be varying
 instead. This conclusion is also supported by the abnormal
 $\chi^2_{min}=1281.5972$ of the $y$-0 case, which is significantly
 larger than the ones of the $y$-1, $y$-2 cases
 (see Tabel~\ref{tab5}). In both the $y$-1, $y$-2 cases,
 $q_0<0$ beyond $3\sigma$ C.L., and this means that
 the universe is undergoing an acceleration.
 On the other hand, $\gamma_0\simeq 0.42$ is inconsistent with
 the latest observational data far beyond $3\sigma$ C.L. in
 both the $y$-1, $y$-2 cases. $\gamma_0\simeq 0.55$ is well
 consistent with the latest observational data within $1\sigma$
 region in the $y$-1 case, but it is inconsistent with the
 latest observational data beyond $3\sigma$ C.L. in the $y$-2
 case. A varying $\gamma$ with non-zero $\gamma_1$ and/or
 $\gamma_2$ is favored. It is easy to see that $\gamma_1<0$ far
 beyond $3\sigma$ C.L. in both the $y$-1, $y$-2 cases, and
 $\gamma_2>0$ far beyond $3\sigma$ C.L. in the $y$-2 case. However,
 $\gamma=\gamma(y)=\gamma(z/(1+z))$, and hence one should be
 careful to treat $\gamma$ as a function of redshift $z$.


\begin{table}[tb]
 \renewcommand{\arraystretch}{1.6}
 \begin{center}
 \vspace{-2mm}  
 \begin{tabular}{c|c|c|c} \hline\hline
  Parameters \ &  Case $y$-0  &  Case $y$-1 & Case $y$-2 \\ \hline
  $\Omega_{m0}$ & \ $0.3713_{-0.0165 }^{+0.0165}{}_{-0.0316}^{+0.0333}{}_{-0.0411}^{+0.0443}$ \
     & \ $0.3482_{-0.0164 }^{+0.0148}{}_{-0.0286}^{+0.0299}{}_{-0.0370}^{+0.0396}$ \
     & \ \ $0.3624_{-0.0174 }^{+0.0140}{}_{-0.0302}^{+0.0323}{}_{-0.0327}^{+0.0428}$  \\
  $q_0$ & \ $0.4819_{-0.1728 }^{+0.1733}{}_{-0.3394}^{+0.3422}{}_{-0.4527}^{+0.4482}$ \
     & \ \ $-0.5427_{-0.1968 }^{+0.1773}{}_{-0.3189}^{+0.3558}{}_{-0.4759}^{+0.4642}$ \ \
     & \ \ $-0.6581_{-0.1858 }^{+0.1771}{}_{-0.3646}^{+0.3486}{}_{-0.4431}^{+0.4694}$  \\
  $j_0$ & \ \ $-9.0694_{-1.1371 }^{+0.8780}{}_{-1.9338}^{+2.0481}{}_{-2.2762}^{+3.0330}$ \ \
     & \ $0.3346_{-1.9620}^{+1.9287}{}_{-4.0999}^{+4.0175}{}_{-5.1470}^{+5.7050}$ \
     & \ \ $1.8465_{-2.6193 }^{+2.2358}{}_{-4.3478}^{+4.6914}{}_{-5.5699}^{+5.8584}$  \\
  $\sigma_{8,\,0}$ & \ $0.7134_{-0.0178 }^{+0.0178}{}_{-0.0345}^{+0.0357}{}_{-0.0450}^{+0.0474}$ \
     & \ $0.7397_{-0.0164 }^{+0.0162}{}_{-0.0329}^{+0.0332}{}_{-0.0422}^{+0.0473}$ \
     & \ \ $0.7224_{-0.0158 }^{+0.0157}{}_{-0.0307}^{+0.0293}{}_{-0.0390}^{+0.0397}$  \\
  $\gamma_{0}$ & \ $0.2443_{-0.0345 }^{+0.0386}{}_{-0.0753}^{+0.0693}{}_{-0.1020}^{+0.0900}$ \
     & \ $0.5694_{-0.0405 }^{+0.0406}{}_{-0.0830}^{+0.0820}{}_{-0.1052}^{+0.1069}$ \
     & \ \ $0.7444_{-0.0689 }^{+0.0645}{}_{-0.1430}^{+0.1605}{}_{-0.1713}^{+0.1935}$  \\
  $\gamma_{1}$ & N/A & \ $-1.0105_{-0.0758 }^{+0.0749}{}_{-0.1437}^{+0.1502}{}_{-0.1873}^{+0.1920}$ \
     & \ \ $-2.2070_{-0.3524 }^{+0.4130}{}_{-0.7170}^{+0.6597}{}_{-1.0784}^{+0.8570}$  \\
  $\gamma_{2}$ & N/A & N/A & \ \ $1.5351_{-0.4941 }^{+0.4414}{}_{-0.8149}^{+0.8089}{}_{-1.1093}^{+1.4499}$  \\
 \hline\hline
 \end{tabular}
 \end{center}
 \caption{\label{tab3} The mean with $1\sigma$, $2\sigma$,
 $3\sigma$ marginalized uncertainties of the model parameters
 for the cases with the $y$-cosmography and $\gamma=\gamma_0$
 (labeled as ``\,$y$-0\,''), $\gamma=\gamma_0+\gamma_1\,y$ (labeled
 as ``\,$y$-1\,''), $\gamma=\gamma_0+\gamma_1\,y+\gamma_2\,y^2$
 (labeled as ``\,$y$-2\,''). See the text for details.}
 \vspace{-3mm}  
 \end{table}



\vspace{-11mm}  


\section{Observational constraints with the Pad\'e cosmography}\label{sec4}

In the previous section, two types of ordinary cosmography are
 considered. As mentioned above, the $z$-cosmography might
 diverge at high redshift $z$. So, the $y$-cosmography was
 proposed as an alternative in the literature, which converges
 in the whole cosmic past $0\leq z<\infty$. However, there
 still exist several problems in the $y$-cosmography. In
 practice, the Taylor series should be truncated by throwing
 away the higher order terms, since it is difficult to deal
 with infinite series. So, the error of a Taylor approximation
 with lower order terms will become unacceptably large when
 $y=z/(1+z)$ is close to $1$ (say, when $z>9$). On the other
 hand, the $y$-cosmography cannot work well in the cosmic
 future $-1<z<0$. The Taylor series with respect to $y=z/(1+z)$
 does not converge when $y<-1$ (namely $z<-1/2$), and it
 drastically diverges when $z\to -1$ (it is easy to see that
 $y\to -\infty$ in this case). Therefore, in~\cite{Zhou:2016nik}, we
 proposed some generalizations of cosmography inspired by the
 Pad\'e approximant, which can avoid or at least alleviate the
 problems of ordinary cosmography.

The so-called Pad\'e approximant can be regarded as a generalization
 of the Taylor series. For any function $F(x)$, its Pad\'e
 approximant of order $(m,\,n)$ is given by the rational
 function~\cite{Pade1892,Padewiki,PadeTalk} (see
 also e.g.~\cite{Adachi:2011vu,Gruber:2013wua,Wei:2013jya,
 Liu:2014vda,Capozziello:2018jya,Aviles:2014rma,Capozziello:2017ddd})
 \be{eq16}
 F(x)=\frac{\alpha_0+\alpha_1 x+\cdots+\alpha_m x^m}{1+
 \beta_1 x+\cdots+\beta_n x^n}\,,
 \ee
 where $m$ and $n$ are both non-negative integers, and
 $\alpha_i$, $\beta_i$ are all constants. Obviously, it
 reduces to the Taylor series when all $\beta_i=0$. Actually in
 mathematics, a Pad\'e approximant is the best approximation of
 a function by a rational function of given
 order~\cite{Padewiki}. In fact, the Pad\'e approximant often
 gives a better approximation of the function than truncating
 its Taylor series, and it may still work where the Taylor
 series does not converge~\cite{Padewiki}.

One can directly parameterize the dimensionless luminosity
 distance based on the Pad\'e approximant with respect to
 redshift $z$~\cite{Zhou:2016nik},
 \be{eq17}
 D_L=\frac{H_0 d_L}{c}=\frac{\alpha_0+\alpha_1 z
 +\cdots+\alpha_m z^m}{1+\beta_1 z+\cdots+\beta_n z^n}\,.
 \ee
 Following~\cite{Zhou:2016nik}, we consider a moderate order
 $(2,\,2)$ in this work, and then
 \be{eq18}
 D_L\equiv\frac{H_0 d_L}{c}=\frac{\alpha_0+\alpha_1 z+\alpha_2 z^2}
 {1+\beta_1 z+\beta_2 z^2}\,.
 \ee
 Obviously, it can work well in the whole redshift range
 $-1<z<\infty$, including not only the past but also the future
 of the universe. In particular, it is still finite even when
 $z\gg 1$. In fact, this $D_L$ was confronted with Union2.1
 SNIa data and Planck 2015 CMB data in~\cite{Zhou:2016nik},
 and the parameters $\alpha_0$ and $\beta_2$ were found to be
 very close to $0$ even in the $3\sigma$ region. So, in the
 present work, it is safe to directly set
 \be{eq19}
 \alpha_0=\beta_2=0\,,
 \ee
 and then the free parameters are now $\alpha_1$, $\alpha_2$
 and $\beta_1$. Note that in fact $\alpha_0=0$ is required by
 $d_L(z=0)=0$ theoretically. On the other hand, we can also
 expand $\gamma(z)$ as a Taylor series with respect to redshift $z$,
 namely $\gamma(z)=\gamma_0+\gamma_1\,z+\gamma_2\,z^2+...$.
 Again, we consider three cases, labeled as ``\,P-0\,'',
 ``\,P-1\,'', ``\,P-2\,'', in which $\gamma(z)$ is Taylor
 expanded up to zeroth, first, second orders, respectively.


\begin{table}[tb]
 \renewcommand{\arraystretch}{1.6}
 \begin{center}
 \vspace{-1mm}  
 \begin{tabular}{c|c|c|c} \hline\hline
  Parameters \ &  Case P-0  &  Case P-1 & Case P-2 \\ \hline
  $\Omega_{m0}$ & \ \ $0.3277_{-0.0210 }^{+0.0210}{}_{-0.0403}^{+0.0428}{}_{-0.0523}^{+0.0570}$ \ \
     & \ $0.3366_{-0.0344 }^{+0.0275}{}_{-0.0604}^{+0.0637}{}_{-0.0727}^{+0.0929}$ \
     & \ \ $0.3407_{-0.0351 }^{+0.0291}{}_{-0.0623}^{+0.0647}{}_{-0.0750}^{+0.0911}$  \\
  $\alpha_1$ & \ $0.9471_{-0.0329 }^{+0.0328}{}_{-0.0631}^{+0.0670}{}_{-0.0820}^{+0.0899}$ \
     & \ $0.9362_{-0.0458 }^{+0.0458}{}_{-0.0909}^{+0.0904}{}_{-0.1194}^{+0.1197}$ \
     & \ \ $0.9307_{-0.0458 }^{+0.0458}{}_{-0.0883}^{+0.0916}{}_{-0.1141}^{+0.1219}$  \\
  $\alpha_2$ & \ $1.0216_{-0.0442 }^{+0.0442}{}_{-0.0863}^{+0.0880}{}_{-0.1129}^{+0.1163}$ \
     & \ $1.0083_{-0.0598 }^{+0.0600}{}_{-0.1212}^{+0.1146}{}_{-0.1605}^{+0.1480}$ \
     & \ \ $1.0012_{-0.0581 }^{+0.0581}{}_{-0.1147}^{+0.1154}{}_{-0.1493}^{+0.1517}$  \\
  $\beta_1$ & \ $0.3329_{-0.0153 }^{+0.0153}{}_{-0.0293}^{+0.0309}{}_{-0.0381}^{+0.0412}$ \
     & \ $0.3324_{-0.0161 }^{+0.0148}{}_{-0.0295}^{+0.0313}{}_{-0.0384}^{+0.0415}$ \
     & \ \ $0.3321_{-0.0154 }^{+0.0154}{}_{-0.0297}^{+0.0312}{}_{-0.0385}^{+0.0418}$  \\
  $\sigma_{8,\,0}$ & \ $0.7615_{-0.0245 }^{+0.0220}{}_{-0.0440}^{+0.0478}{}_{-0.0566}^{+0.0644}$ \
     & \ $0.7526_{-0.0337 }^{+0.0338}{}_{-0.0662}^{+0.0684}{}_{-0.0864}^{+0.0921}$ \
     & \ \ $0.7480_{-0.0362 }^{+0.0329}{}_{-0.0658}^{+0.0702}{}_{-0.0840}^{+0.0937}$  \\
  $\gamma_0$ & \ $0.5462_{-0.0539 }^{+0.0538}{}_{-0.1038}^{+0.1089}{}_{-0.1343}^{+0.1454}$ \
     & \ $0.5447_{-0.0544 }^{+0.0543}{}_{-0.1051}^{+0.1097}{}_{-0.1374}^{+0.1470}$ \
     & \ \ $0.5562_{-0.0590 }^{+0.0590}{}_{-0.1151}^{+0.1187}{}_{-0.1511}^{+0.1583}$  \\
  $\gamma_1$ &  N/A  & \ \ $-0.0667_{-0.1908 }^{+0.2110}{}_{-0.4122}^{+0.3801}{}_{-0.5546}^{+0.4923}$ \ \
     & \ \ $-0.2488_{-0.3854 }^{+0.3858}{}_{-0.7665}^{+0.7622}{}_{-1.0127}^{+1.0016}$  \\
  $\gamma_2$ & N/A & N/A & \ \ $0.2378_{-0.4540 }^{+0.4531}{}_{-0.9190}^{+0.8788}{}_{-1.2341}^{+1.1400}$  \\
 \hline\hline
 \end{tabular}
 \end{center}
 \caption{\label{tab4} The mean with $1\sigma$, $2\sigma$,
 $3\sigma$ marginalized uncertainties of the model parameters
 for the cases with the Pad\'e cosmography and
 $\gamma=\gamma_0$ (labeled as ``\,P-0\,''),
 $\gamma=\gamma_0+\gamma_1\,z$ (labeled as ``\,P-1\,''),
 $\gamma=\gamma_0+\gamma_1\,z+\gamma_2\,z^2$ (labeled as
 ``\,P-2\,''). See the text for details.}
 \end{table}



 \begin{center}
 \begin{figure}[tb]
 \centering
 \vspace{-12mm}  
 \includegraphics[width=0.7\textwidth]{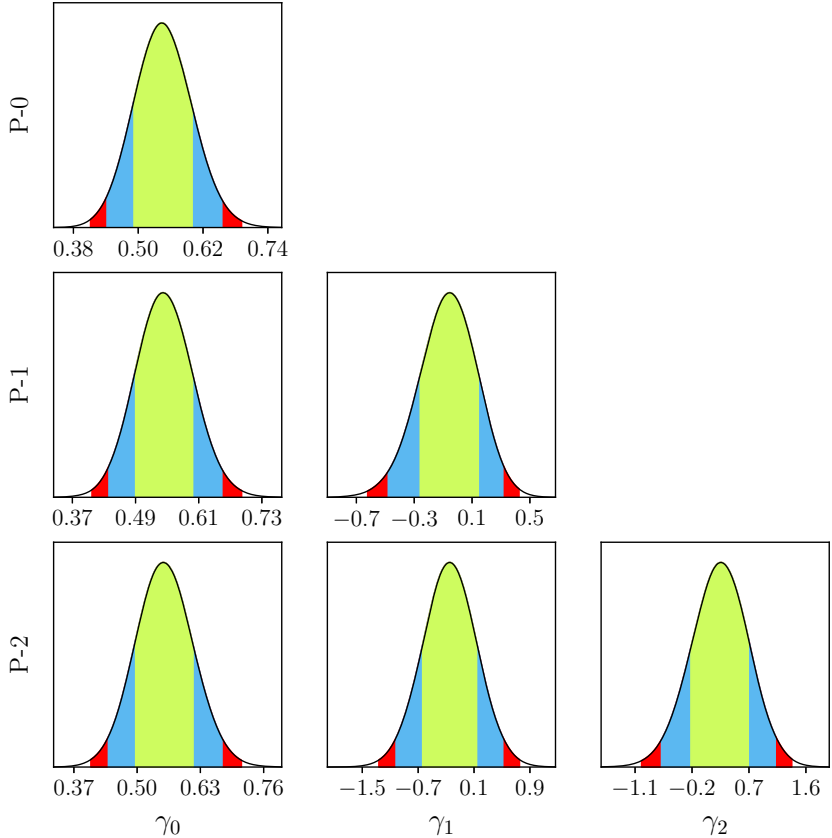}
 \caption{\label{fig3} The same as in Fig.~\ref{fig1}, except
 for the P-0, P-1, P-2 cases. See the text
 and Table~\ref{tab4} for details.}
 \end{figure}
 \end{center}


\vspace{-10.8mm}  

Since the Pad\'e cosmography still works well at very high
 redshift $z\gg 1$ in contrast to the ordinary cosmography as
 mentioned above, in this section, we further use the latest
 CMB data in addition to the observational data mentioned in
 Sec.~\ref{sec2}. However, using the full data of CMB to
 perform a global fitting consumes a large amount
 of computation time and power. As a good alternative, one can
 instead use the shift parameter $R$~\cite{Bond:1997wr} from
 CMB, which has been used extensively in the literature
 (including the works of the Planck and the WMAP
 Collaborations themselves). It is argued in e.g.~\cite{Wang:2006ts,
 Wang:2013mha,Shafer:2013pxa} that the shift parameter $R$ is
 model-independent and contains the main information of the
 observation of CMB. As is well known, the shift parameter $R$
 is defined by~\cite{Bond:1997wr,Wang:2006ts,Wang:2013mha,
 Shafer:2013pxa}
 \be{eq20}
 R\equiv\sqrt{\Omega_{m0}H^2_0}\,\left(1+z_\ast\right)d_A(z_\ast)/c
 =\frac{\sqrt{\Omega_{m0}}\, D_L(z_\ast)}{1+z_\ast}\,,
 \ee
 where the redshift of the recombination $z_\ast=1089.92$ from
 the Planck 2018 data~\cite{Aghanim:2018eyx}, and the angular
 diameter distance $d_A$ is related to the luminosity distance
 $d_L$ through $d_A=d_L(1+z)^{-2}$ (see e.g. the
 textbooks~\cite{Weinberg1972}). Here we adopt the value
 $R_{obs}=1.7502\pm 0.0046$~\cite{Chen:2018dbv} derived from
 the Planck 2018 data. Thus, the corresponding $\chi^2$ from
 the latest CMB data is given by
 $\chi^2_R=(R_{\rm mod}-R_{obs})^2/\sigma_R^2$. Although the
 number of data points $\cal N$ and the number of free
 parameters $\kappa$ both increase by $1$, the degree of
 freedom $dof={\cal N}-\kappa$ is unchanged in this case. It is
 worth noting that the acoustic scale $l_A$, and $\Omega_b h^2$, the
 scalar spectral index $n_s$ are commonly used with the shift
 parameter $R$ in the literature, but they will introduce extra
 model parameters as mentioned above, and hence we do not use
 them here.

By using the latest observational data, we obtain
 the constraints on all the model parameters involved, and
 present them in Table~\ref{tab4}, for the P-0, P-1, P-2
 cases. In Fig.~\ref{fig3}, we also present the 1D marginalized
 probability distributions of the parameters related to the
 growth index $\gamma$, namely $\gamma_0$, $\gamma_1$ and
 $\gamma_2$. In all cases, $\gamma_0\simeq 0.42$ is inconsistent
 with the latest observational data beyond $2\sigma$ C.L. (but
 it could be consistent in the $3\sigma$ region). On the other
 hand, in all cases, $\gamma_0\simeq 0.55$ is well consistent
 with the latest observational data within the $1\sigma$
 region. Note that in all cases, a constant $\gamma=\gamma_0$
 (namely $\gamma_1=0$ and $\gamma_2=0$) is well consistent with
 the latest observational data within the $1\sigma$ region
 (but see below).


\section{Conclusion and discussion}\label{sec5}

In this work, we consider the constraints on the growth index
 $\gamma$ by using the latest observational data. To be
 model-independent, we use cosmography to describe the cosmic
 expansion history, and also expand the general $\gamma(z)$ as
 a Taylor series with respect to redshift $z$ or $y$-shift,
 $y=1-a=z/(1+z)$. We find that the present value
 $\gamma_0=\gamma(z=0)\simeq 0.42$ (for most of viable $f(R)$
 theories) is inconsistent with the latest observational data
 beyond $3\sigma$~C.L. in the 6 cases with the
 usual cosmography, or beyond $2\sigma$~C.L. in the 3 cases
 with the Pad\'e cosmography. This result supports our previous
 work~\cite{Yin:2018mvu}. On the other hand,
 $\gamma_0\simeq 0.55$ (for dark energy models in GR) is consistent
 with the latest observational data at $1\sigma$ C.L. in 5 of the 9
 cases under consideration, but is inconsistent beyond $2\sigma$ C.L.
 in the other 4 cases (while it is still consistent within the
 $3\sigma$ region). Therefore, we can say nothing firmly about
 $\gamma_0\simeq 0.55$. This result is still consistent with
 the reconstructed $\gamma(z)$ at $z=0$ obtained in our
 previous work~\cite{Yin:2018mvu}. A varying $\gamma$ with non-zero
 $\gamma_1$ and/or $\gamma_2$ is favored in the cases with the usual
 cosmography, while in the cases with the Pad\'e cosmography, a
 constant $\gamma=\gamma_0$ (namely $\gamma_1=0$ and $\gamma_2=0$) can
 still be consistent with the latest observational data
 (but this might be artificial, see below).

\begin{center}
 \begin{figure}[tb]
 \centering
 \vspace{-9mm}  
 \includegraphics[width=0.45\textwidth]{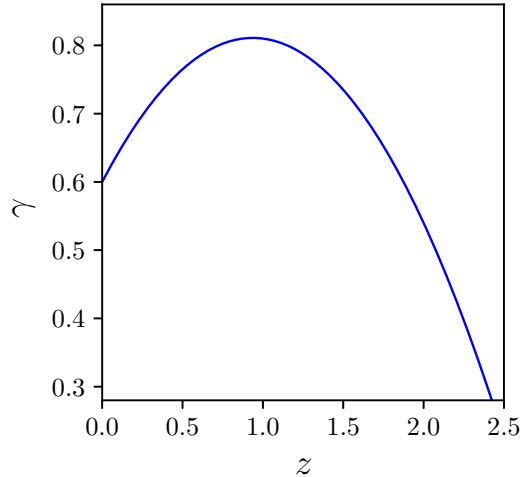}
 \caption{\label{fig4} A demonstration of
 $\gamma=\gamma_0+\gamma_1\,z+\gamma_2\,z^2$ with
 $\gamma_0=0.6$, $\gamma_1=0.45$, $\gamma_2=-0.24$, which are
 all well within the $1\sigma$ regions of their observational
 constraints for the $z$-2 case (see Tabel~\ref{tab2} and
 Fig.~\ref{fig1}).}
 \end{figure}
 \end{center}

It is of interest to compare the 9 cases considered here. We
 adopt several goodness-of-fit criteria used extensively in
 the literature to this end, such as $\chi^2_{min}/dof$,
 $P(\chi^2>\chi^2_{min})$ (see
 e.g.~\cite{Wei:2006ut,Wei:2007ws}), Bayesian Information
 Criterion~(BIC)~\cite{Schwarz:1978} and Akaike Information
 Criterion~(AIC)~\cite{Akaike:1974}, where the degree of
 freedom $dof={\cal N}-\kappa$, while $\cal N$ and $\kappa$
 are the number of data points and the number of free model
 parameters, respectively. The BIC is defined by~\cite{Schwarz:1978}
 \be{eq21}
 {\rm BIC}=-2\ln{\cal L}_{max}+\kappa\ln {\cal N}\,,
 \ee
 and the AIC is defined by~\cite{Akaike:1974}
 \be{eq22}
 {\rm AIC}=-2\ln{\cal L}_{max}+2\kappa\,,
 \ee
 where ${\cal L}_{max}$ is the maximum likelihood. In the
 Gaussian cases, $\chi^2_{min}=-2\ln{\cal L}_{max}$. The
 difference in BIC or AIC between two models makes sense. We
 choose the P-0 case to be the fiducial model when we calculate
 $\Delta$BIC and $\Delta$AIC. In Table~\ref{tab5}, we present
 $\chi^2_{min}/dof$, $P(\chi^2>\chi^2_{min})$, $\Delta$BIC and
 $\Delta$AIC for the 9 cases considered in this work. Clearly,
 the cases with $y$-cosmography are the worst, while the cases
 P-0 and $z$-2 are the best. In fact, the goodness-of-fit
 criteria for the cases P-0 and $z$-2 are fairly close. A
 caution should be mentioned here. All the criteria given in
 Table~\ref{tab5} are based on $\chi^2_{min}$, which are read
 from the output .likestats files of the CosmoMC program
 GetDist. However, as the CosmoMC~\cite{Lewis:2002ah} readme
 file puts it, ``\,file$_{-}$root.likestats gives the best fit
 sample model, its likelihood, and ... Note that MCMC does not
 generally provide accurate values for the best-fit.'' Keeping
 this in mind, we could say that the cases P-0 and $z$-2 are
 equally good, since their not so accurate $\chi^2_{min}$ are
 very close actually. In the P-0 case, the growth index
 $\gamma=\gamma_0$ is constant. However, in the $z$-2 case,
 $\gamma_2<0$ beyond $3\sigma$ C.L. (see Table~\ref{tab2} and
 Fig.~\ref{fig1}), and hence the function $\gamma(z)$ is a
 parabola opening down, namely $\gamma$ increases and then
 decreases as redshift $z$ increases. There exists an arched
 structure in the moderate redshift range. This result is quite
 similar to the one of~\cite{Yin:2018mvu}. In Fig.~\ref{fig4}, we show
 a demonstration of $\gamma=\gamma_0+\gamma_1\,z+\gamma_2\,z^2$ with
 $\gamma_0=0.6$, $\gamma_1=0.45$, $\gamma_2=-0.24$, which are
 all well within the $1\sigma$ regions of their observational
 constraints for the $z$-2 case (see Tabel~\ref{tab2} and
 Fig.~\ref{fig1}).


 \begin{table}[tb]
 \renewcommand{\arraystretch}{1.6}
 \begin{center}
 \resizebox{\textwidth}{!}{ 
 \begin{tabular}{llllllllll} \hline\hline
  Cases  & $z$-0 & $z$-1 & $z$-2 & $y$-0 & $y$-1 & $y$-2 & P-0 & P-1 & P-2  \\ \hline
  $\chi^2_{min}$  & 1106.2202 \ & 1101.3110 \ & 1093.0770 \ & 1281.5972 \ & 1129.5110 \ & 1116.2714 \ & 1094.7438 \ & 1094.7268 \ & 1094.3310 \\
  $\cal N$ & 1127 & 1127 & 1127 & 1127 & 1127 & 1127 & 1128 & 1128 & 1128 \\
  $\kappa$ & 5 & 6 & 7 & 5 & 6 & 7 & 6 & 7 & 8 \\
  $\chi^2_{min}/dof$  & 0.9859 & 0.9824 & 0.9760 & 1.1422 & 1.0076 & 0.9967 & 0.9757 & 0.9766 & 0.9771 \\
  $P(\chi^2 > \chi^2_{min})$ \quad  & 0.6257 & 0.6570 & 0.7120 & 0.0006 & 0.4233 & 0.5258 & 0.7143 & 0.7072 & 0.7028 \\
  $\Delta {\rm BIC}$  &  4.4438 & 6.5619 & 5.3552 & 179.8210 & 34.7619 & 28.5496 & 0 & 7.0112 & 13.6436 \\
  $\Delta {\rm AIC}$  &  9.4764 & 6.5672 & 0.3332 & 184.8530 & 34.7672 & 23.5276 & 0 & 1.9830 & 3.5872 \\
  Rank  &  6 & 5 & 2 & 9 & 8 & 7 & 1 & 3 & 4  \\
 \hline\hline
 \end{tabular}
 } 
 \end{center}
 \caption{\label{tab5} Comparing the 9 cases considered in
 the present work. See the text for details and caution.}
 \end{table}


It is worth noting that throughout this work, we always consider the
 growth index $\gamma$ as a Taylor series with respect to $z$ or $y$,
 namely $\gamma(z)=\gamma_0+\gamma_1\,z+\gamma_2\,z^2+...$, or
 $\gamma(y)=\gamma_0+\gamma_1\,y+\gamma_2\,y^2+...$ However,
 in Sec.~\ref{sec4}, we parameterize the dimensionless
 luminosity distance $D_L$ by using the Pad\'e approximant,
 and hence it can still work well at very high redshift $z\sim 1090$.
 Obviously, it is better to also parameterize the growth index
 $\gamma(z)$ by using the Pad\'e approximant (we thank the
 referee for pointing out this issue). But the cost is
 expensive to do this. If we want to catch the arched structure
 in $\gamma(z)$, at least a Pad\'e approximant of order
 $(2,\,2)$ is needed, which has 5 free parameters (n.b.
 Eq.~(\ref{eq16})), and almost double the number of free parameters
 in a 2nd order Taylor series. So, in the P-2 case, the
 total number of free model parameters will be 10. It will
 consume significantly more computation power and time, but the
 corresponding constraints will be very loose. Therefore, we
 choose not to do this at a great cost. But one should be aware
 of the possible artificial results from this choice. For
 example, $\gamma(z)=\gamma_0+\gamma_1\,z+\gamma_2\,z^2$ will
 diverge at $z\sim 1090$, and hence the values of $\gamma_1$
 and $\gamma_2$ tend to be zero to fit the high-$z$ CMB data
 in the P-1, P-2 cases (we thank the referee for pointing
 out this issue).

Some remarks are in order. First, the growth rate $f$ and then
 the growth index $\gamma$ for modified gravity scenarios
 (especially $f(R)$ theories) in principle are not only
 time-dependent but also scale-dependent (see
 e.g.~\cite{Gannouji:2008wt,Tsujikawa:2009ku}). However, as is
 shown in e.g.~\cite{Gannouji:2008wt,Tsujikawa:2009ku}, the
 behavior of $\gamma$ is nearly scale-independent at low
 redshift $z\,\lsim\,1$, and $\gamma_0=\gamma(z=0)$ is also
 nearly independent of scale. So, this issue does not change
 the main conclusions of the present work, although it may
 be studied carefully in the future work.~Second, as is mentioned in
 the beginning of Sec.~\ref{sec2}, there exist other two approaches
 dealing with the growth history, which numerically solve the
 perturbation equations by using the code CAMB integrated in
 CosmoMC. We will also consider these approaches in the future
 work. Third, in the present work, we do not use some types of
 observational data (for example, the observational $H(z)$
 data, and other kinds of BAO data) to avoid introducing extra
 model parameters. However, in principle, it is not terrible to
 do so, although the constraints might be loose and the calculations
 might be complicated. Finally, in this work, we only consider
 the Taylor series expansion of the growth index $\gamma$ up to
 2nd order, and the usual cosmography up to 3rd order. In fact,
 one can also further consider higher orders in these cases. We
 anticipate that our main conclusions will not change significantly.


\vspace{-10mm}  


\section*{ACKNOWLEDGEMENTS}

We thank the anonymous referee for useful comments and suggestions,
 which helped us to improve this work. We are grateful to
 Hua-Kai~Deng, Da-Chun~Qiang, Xiao-Bo~Zou, Zhong-Xi~Yu,
 and Shou-Long~Li for kind help and discussions. This work
 was supported in part by NSFC under Grants No.~11575022
 and No.~11175016.

\renewcommand{\baselinestretch}{1.0}


\end{document}